\documentclass[useAMS]{mn2e}

\usepackage{amssymb,amsmath,psfig,times}
\def\gsim{ \lower .75ex \hbox{$\sim$} \llap{\raise .27ex \hbox{$>$}} }
\def\lsim{ \lower .75ex\hbox{$\sim$} \llap{\raise .27ex \hbox{$<$}} }
\def\crexp{{\rm\thinspace km^{2} \thinspace sr \thinspace yr}}

\def\ev{{\rm\thinspace eV}}

\def\hi{{\rm H}\,{\small\rm I}}

\title[Sources of UHECRs]
{Ultra--High Energy Cosmic Rays, Spiral Galaxies and Magnetars}  
\author[G. Ghisellini et al.]
{G. Ghisellini,$^1$\thanks{Email:
gabriele.ghisellini@brera.inaf.it} G. Ghirlanda,$^1$ F. Tavecchio,$^{1}$,
 F. Fraternali$^2$ and G. Pareschi$^1$\\
$^1$INAF -- Osservatorio Astronomico di Brera, Via Bianchi 46 Merate, Italy\\
$^2$Dept. of Astronomy, University of Bologna, via Ranzani 1, 40127 Bologna, Italy 
}

\begin{document}  

\maketitle

\begin{abstract}
We measure the correlation between the arrival directions of the
highest energy cosmic rays detected by the Pierre Auger Observatory
with the position of the galaxies in the \hi\ Parkes All Sky Survey
(HIPASS) catalogue, weighted for their \hi\ flux and Auger
exposure. The use of this absorption--free catalogue, complete also
along the galactic plane, allows us to use all the Auger events. The
correlation is significant, being 86.2\% for the entire 
sample of \hi\ galaxies, and becoming 99\% when considering the richest 
galaxies in HI content, or 98\% with those lying between 40--55 Mpc. 
We interpret this result as the evidence that spiral galaxies are the
hosts of the producers of UHECR and we briefly discuss classical (i.e
energetic and distant) long Gamma Ray Burst (GRBs), short GRBs, as
well as newly born or late flaring magnetars as possible sources of
the Auger events. With the caveat that these events are still very
few, and that the theoretical uncertainties are conspicuous, we
found that newly born magnetars are the best candidates.  If so,
they could also be associated with sub--energetic, spectrally soft,
nearby, long GRBs. 
We finally discuss why there is a clustering of Auger events in 
the direction on the 
radio--galaxy Cen A and an absence of events in the direction of the
radio--galaxy M87.
\end{abstract}


\section{Introduction}

The origin of ultra--high energy cosmic rays (UHECR), exceeding 10 EeV
(1 EeV=$10^{18}$ eV) has been a mystery for decades, but the recent
findings of the large area detectors, such as AGASA (Ohoka et al. 1997), HIRes
(Abu--Zayyad et al. 2000), and especially the Pierre Auger Southern Observatory
(Abraham et al. 2004), began to disclose crucial clues about the
association of the highest energy events with cosmic sources.  
The Auger collaboration (Abraham et al. 2007) found a
positive correlation between the arrival directions of UHECR with
energies grater than 57 EeV and nearby AGNs (in the 
optical catalogue of Veron--Cetty \& Veron 2006).  
Although this result has not been
confirmed by HIRes (Abbasi et al. 2008) and it has been criticised by
Gorbunov et al. (2008), it received an important confirmation by
George et al.  (2008), who considered a complete sample of nearby hard
X--ray emitting AGNs detected by the BAT instrument onboard {\it Swift}.  
This sample is much less affected by absorption than any
optical sample although, to identify as such an AGN, one relies on
optical identification.  
Moreover, George et al. (2008) found a correlation
not simply with the AGN locations, but by weighting them for
the X--ray flux and the Auger exposure.

This association, if real, is surprising, since the large majority of
the correlating AGNs are radio--quiet, a class of objects not showing,
in their electromagnetic spectrum, any sign of non--thermal high
energy emission: no radio--quiet AGN was detected by the EGRET
instrument onboard the {\it Compton Gamma Ray Observatory} (Hartman et
al. 1999).  Therefore they must accelerate particles (protons, nuclei,
and presumably their accompanying electrons) to ultra--high energies
without any noticeable radiative emission from these very same
particles.  Radio--loud AGNs, instead, together with Gamma Ray Bursts
(of both the long and short category) do show high energy non--thermal
emission, and have been considered for a long time better candidates
as UHECR sources (Vietri 1995; Waxman 1995; Milgrom \& Usov 1995;
Wang, Razzaque \& M\'esz\'aros 2008; 
Murase et al. 2008; 
Torres \& Anchordoqui 2004 
and Dermer 2007 for reviews, 
and Nagar \& Matulich 2008
and  Moskalenko et al. 2008 for the possible association of the AUGER
events with radio--loud AGNs). 
Note also that some short GRBs could be
due to the giant flares of highly magnetised neutron stars (``magnetars'', 
as the 27 Dec 2004 event from
1806--20; Borkowski et al. 2004; Hurley et al. 2005; Terasawa et
al. 2005), and that, at birth, a fastly spinning magnetar can be much
more energetic than when, later, it produces giant flares (Arons 2003).

The possibility that GRBs and magnetars are the sites of production of
UHECR would directly imply the direct association of these events with
(normal) galaxies. In this case the found association of UHECR with
nearby AGNs might then be due to the fact that local AGNs just trace
the distribution of galaxies. The aim of the present paper is to test
this possibility directly correlating the locations of the ultra--high
energy Auger events with a well defined, complete, and possibly
absorption--free sample of galaxies.  For this purpose we use the
sample of \hi\ emitting galaxies, compiled using the Parkes 64--m radio
telescope (Barnes et al.  2001; Staveley--Smith et al. 1996), which is
conveniently located in the south hemisphere, as the Auger
observatory.  The entire sample covers the portion of the sky visible
by Auger, making it possible to use, for the correlation analysis, all
the 27 UHECR events with energies larger than 57 EeV detected by
Auger, without excluding the galactic plane, as is instead necessary
when dealing with AGNs or optically selected galaxies. Note that the
presence of neutral hydrogen strongly favours spirals (or, more
generally, gas--rich galaxies) with respect to elliptical galaxies.

We use a cosmology with $\rm{h}_0=\Omega_\Lambda=0.7$ and $\Omega_{\rm M}=0.3$.

\section{Data}

\subsection{UHECR events}

The Auger Observatory (Abraham et al. 2004, 2008), operating in Argentina since
2004, is located at latitude $-35.2\degr$ and it has a maximum zenith angle
acceptance of $60\degr$.  The relative exposure is independent of the energy
of the detected events and it is a nearly uniform in right ascension. The
dependence on declination is given by Sommers (2001). The Observatory can
detect Cosmic Rays from sources with declination $\delta<24.8^\circ$.

\begin{figure}
\vskip -0.5cm
{\centerline {\psfig{figure=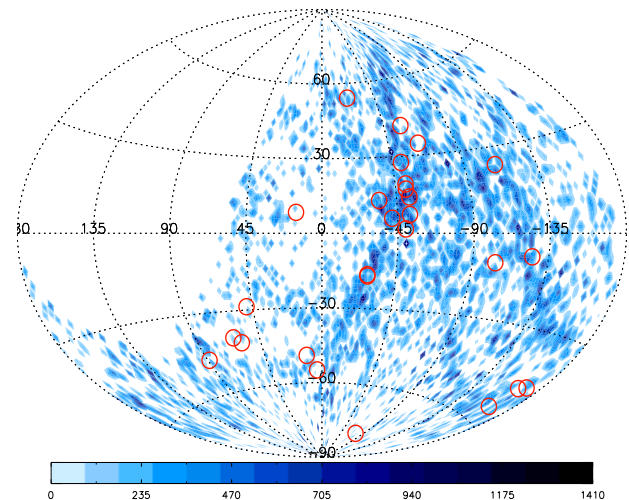,width=9cm,height=7cm}}}
\caption{
HIPASS galaxy \hi\ flux, in galactic coordinates.
Blue levels corresponds to the integrated flux 
(in bins of $2^\circ\times 2^\circ$ and units of Jy km s$^{-1}$)
of \hi\ emission, multiplied by the relative AUGER exposure.
Red circles are the locations of the 27 AUGER UHECRs above 57 EeV.
}
\label{HIflux}
\end{figure}

The available Auger list of UHECR events (Abraham et al. 2008) comprises 27
events with energies in excess of $5.7\times{}10^{19}\ev$ from an integrated
exposure of $9000\crexp$.  The event arrival directions are determined with an
angular resolution of better than $1\degr$. However, magnetic fields of
unknown strength will deflect charged particles on their trajectories through
space.  The advantage of studying the highest energy events is that this
deflection is minimised, but it can still be up to $\sim{}10\degr$ in the
Galactic field.  
The 27 UHECR detected by Auger are
distributed in the range $\delta\in[-61,9.6]$ (or at galactic latitudes
$b\in[-78.6,54.1]$ -- open circles in Fig. \ref{HIn} and Fig. \ref{HIflux}).

\subsection{HIPASS catalogue}
 
\begin{figure}
\vskip -0.5cm
{\psfig{figure=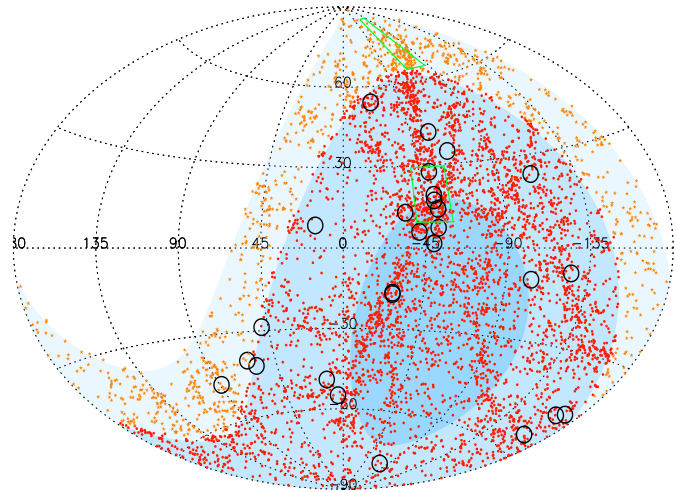,width=9cm,height=7cm}} 
\caption{
\hi\ Parkes All--Sky Survey (HIPASS) galaxies  (solid filled symbols) 
and the 27 AUGER UHECRs (black open circles) in galactic coordinates.  Blue
levels corresponds to the AUGER relative exposure. The south \hi\ catalogue
(HICAT -- 4315 sources) and the north extension (NHICAT -- 1002 sources) are
shown by red filled circles and orange stars, respectively.
The green lines are $20^\circ\times 20^\circ$ boxes centred on 
the positions of the radio--galaxies M87 (near the north Galactic pole) 
and Cen A.
}
\label{HIn}
\end{figure}

We compare the arrival directions of Auger UHECRs with the locations
of sources of the \hi\ Parkes All--Sky Survey (HIPASS -- Meyer et
al. 2004).  This is a blind survey of sources in \hi\ covering the full
southern sky at $\delta<25^\circ$ which is the same sky area accessed
by the Auger Observatory.  The full catalogue is composed by a list of
4315 sources at $\delta<2^\circ$ (HICAT -- Meyer et al. 2004; Zwaan et
al., 2004) and by its extension to the northern sky up to
$\delta=25^\circ$ (NHICAT -- Wong et al. 2006) which includes 1002
sources.  All sources are shown in Fig. \ref{HIn} with the 27 UHECR
detected by Auger.

The HICAT and NHICAT have different level of completeness.   
To have a catalogue complete in flux at the 95\% level, we cut
the HICAT at $S_{\rm int}>7.4$ Jy km s$^{-1}$ and the NHICAT at 
$S_{\rm int}>15$ Jy km s$^{-1}$ as discussed in
Zwaan et al. (2004) and Wong et al. (2006). 
$S_{\rm int}$ represents the total \hi\ line flux.
For the purposes of this paper we also considered the \hi\
sources within 100 Mpc which is the maximum distance at which UHECRs
of $E>57$ EeV can survive the GZK suppression effect (see e.g. Harari et al. 2006).
We call this sample 95HIPASS: 
it contains 2414 sources from the HICAT and 290
sources from the NHICAT for a total of 2704 sources and covers the
entire sky at $\delta<25^\circ$.  We will also consider the southern
sky sample alone which is more complete and can be cut at 99\%
completeness for $S_{\rm int}>9.4$ Jy km s$^{-1}$ (also by considering
sources at $<$100 Mpc). This sample contains 1946 sources and is
called 99HICAT.

\section{Analysis}

To quantify the possible correlation
between UHECR Auger events and the distribution of \hi\ local galaxies 
we use the method adopted by George et al. (2008). In order to quantify
the probability that two sets of sources are drawn from the same parent
population of objects we perform the two-dimensional generalisation of the
Kolmogorov--Smirnov (K--S) test (Peacock 1983) proposed by Fasano \&
Franceschini (1987).

In our case the test is used to compare two data samples, i.e. the UHECR and
the \hi\ galaxies. 
This test can then measure either if UHECRs have a galaxy counterpart, and, 
viceversa, if a concentration of galaxies has an UHECR counterpart.
%
The test relies on the statistic $D$, also used for the
unidimensional K--S test, which represents the maximum difference between the
cumulative distributions of the two data samples. 
For each
UHECR data point $j$ we compute a set of four numbers $d_{j,i}$ ($i$=[1,4])
defined as the difference of the relative fraction of UHECR and \hi\ galaxies
found in the four natural quadrants defined around point $j$. Hence,
$D=\max(d_{j,i})$ for all the data points considered. 
Defining $Z_{\rm {n}}=D\sqrt{n}$,
the strength of the correlation between two catalogues is 
the integral probability distribution $P(D\sqrt{n}>\rm{observed})$, where
$n=N_1N_2/(N_1+N_2)$, and $N_1$ and $N_2$ are the number of data points in the
two sets.  This measurement can be used to determine the similarity of sets of
positions on the sky.  

The probability can be computed analytically for large data sets ($n>$80 --
Fasano \& Franceschini 1987). 
In our case, having only 27 UHECR, we have to
rely on Monte Carlo simulations.  
We generate a large set of random UHECR
events according to the relative Auger exposure. For each synthetic UHECR
sample we compute $Z_{\rm {n}}$ by correlating it with the catalogue
of \hi\ galaxies. The probability of the observed $Z_{\rm {n}}$ is given by the
number of times we find a value of $Z_{\rm {n}}$ larger than the observed one.
This is the probability that the correlation between the (real) UHECR
sample and the \hi\ galaxies is not by chance. Large (low) values of the
probability indicate a good (poor) correlation between the Auger UHECRs and
the given \hi\ galaxy sample.

As noted by George et al. (2008), the two--dimensional K--S test can
be performed with the number of data points or with the flux of the
sources in the comparison sample. 
In our case $D$ represents the maximum difference
between the number of UHECRs and that of the sum of the galaxies weighted for their
flux and for the the relative Auger exposure.  
The advantage of using the weighted flux of the sources is that it accounts 
for their distance.  
George et al. (2008) found that the UHECRs are  more 
correlated with the weighted flux of {\it Swift} AGN than with 
with their position.  
In Fig. \ref{HIflux} we show the map of the flux of the HIPASS catalogue
weighted for the Auger relative exposure.

\begin{figure*}
\centerline{\psfig{figure=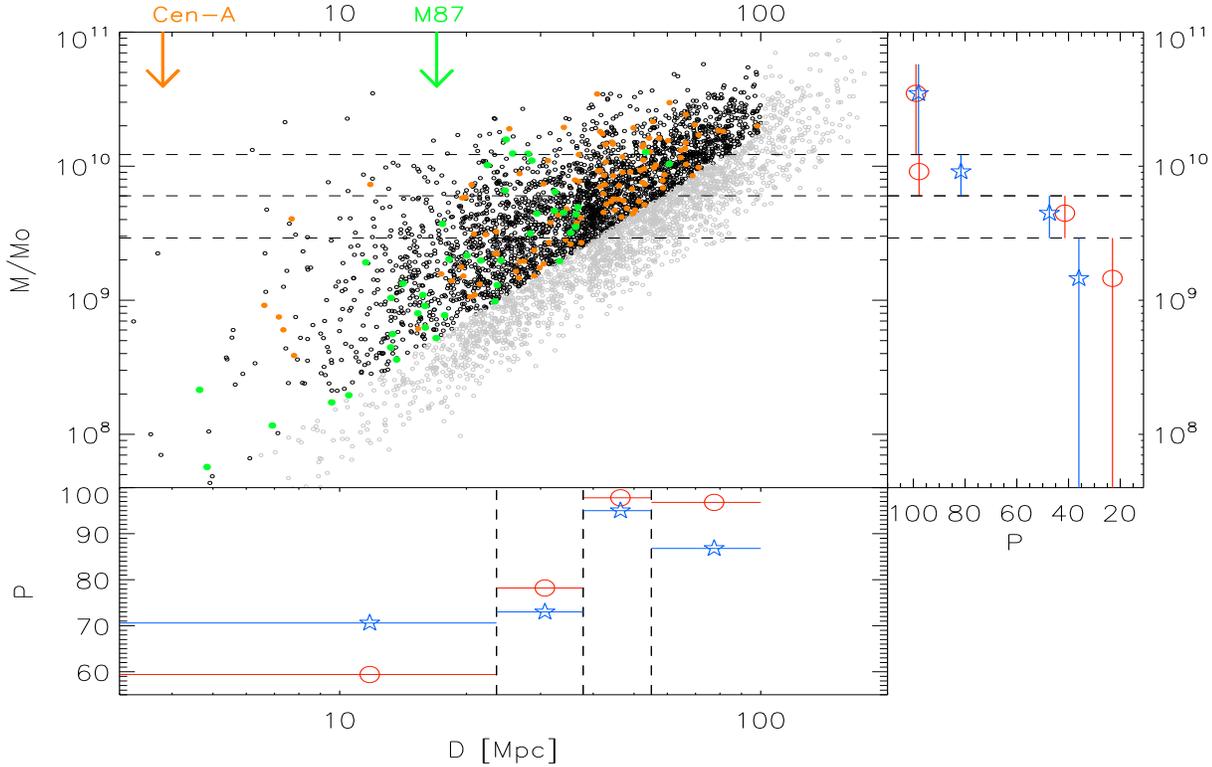,width=17cm,height=11cm}} 
\vskip -0.5 true cm
\caption{
Top left panel: the HI mass (in solar masses)
of our galaxies as a function of their distance.
The \hi\ mass is proportional to the \hi\ luminosity, and is found using
$M=2.36\times 10^5 D_{\rm Mpc}^2 S_{\rm int}$,
where $S_{\rm int}$ is the integrated flux in [Jy km/s].
Black empty circles are those galaxies forming a complete, flux limited, sample.
Orange and green filled circles are galaxies in the $20^\circ \times 20^\circ$ boxes centred
on the position of the radio--galaxies Cen A and M87, respectively.
The distances of these two radio--galaxies are marked by an arrow.
At the distance of Cen A there is almost no \hi\ emitting galaxy, and
no concentration is seen at the distance of M87 (and the Virgo cluster).
The \hi\ galaxies lying in the same region of the sky as Cen A show a concentration
for distances 40--50 Mpc, where the Centaurus cluster of galaxies is.
No concentration in distance is seen for \hi\ galaxies in the direction of Virgo.
The bottom left panel shows the significance of the 2D
K--S test using galaxies in different bin of distances
(circles: South sample HICAT; stars: South+North sample HICAT+NHICAT).
The top right panel shows the significance of the correlations for 
different bins of \hi\ content (or, equivalently for different luminosity bins). 
}
\label{bella}
\end{figure*}

\section{Results}

We found that with the 95HIPASS catalogue (2704 \hi\ sources complete in
flux at 95\%)
the probability that UHECRs are correlated with \hi\ galaxies is 71.6\%
by using the weighted flux of the \hi\ sources. 
Considering the more complete 99HICAT 
(1946 \hi\ sources complete in flux at 99\%) distributed within 100 Mpc and
the 25 UHECRs distributed in the same sky region we find a larger
flux--weighted probability of 87.8\%.  
This probability is slightly smaller
than found with local AGN by George et al. (2008).

However, having a large sample of \hi\ galaxies we can study if the
correlation probability changes by considering different sub--samples
of galaxies selected according to their distance or luminosity. We
have considered 4 bins of distance with an equal number of
sources ($\sim$500) per bin. The correlation probability shows a
maximum of 95\% (97.8\% for the 99HICAT) for sources distributed
between 37.8 and 55 Mpc. We show these results in Fig. \ref{bella}
(open circles and stars in the bottom panel).

Similarly we defined four equally populated luminosity bins,
or, equivalently, four bins of HI mass content, since we can use
$M/M_\odot = 2.36\times 10^5 D_{\rm Mpc}^2 S_{\rm int}$ to estimate
the \hi\ mass (here $S_{\rm int}$ is measured in [Jy km/s]).
We find that
the probability (left panel in Fig. \ref{bella}) is maximised by the
most \hi\ luminous or massive (in \hi) sources (98\% and 99\% for 
the 95HIPASS and 99HICAT
sample, respectively, for $M> 1.1\times 10^{10}\, M_{\odot}$).

%

Selecting those \hi\ galaxies located within two $20^\circ \times
20^\circ$ boxes centred on the radio--galaxies Cen A and M87 (green
boxes in Fig. \ref{HIn}), we can show where they lie in the
luminosity--distance plane in Fig. \ref{bella} (orange and green dots,
respectively).  
While there is no clustering of points at the
distances of Cen A and Virgo, we can see that \hi\ galaxies in the
direction of Cen A do cluster at distances of 40--50 Mpc, where the
Centaurus cluster is.  
This could explain why some UHECR
events {\it appear} to be associated with the radio--galaxy Cen A, and
none with M87: beyond Cen A there is the Centaurus cluster, richer of
\hi\ emitting spirals than the Virgo cluster.
The ratio of the integrated HI fluxes from the two $20^\circ\times20^\circ$
boxes (Virgo/Cen A) is 5.9.
To this, we have to multiply by another factor 3 for the lower Auger exposure
in the direction of Virgo.

The sample has too few galaxies beyond 100 Mpc to test the GZK effect 
(that would be revealed by finding no correlation for these galaxies).

\section{Discussion}

The 27 Auger events above 57 EeV, with a total exposure of $9000\crexp$
correspond to an integrated flux, in CGS units:
\begin{equation}
F_{\rm A}(E>57\, {\rm EeV})\, \sim \, 1.1 \times 10^{-11} \,\,\, 
{\rm erg\,\, cm^{-2}\, s^{-1}}
\label{fa}
\end{equation}
This flux is smaller than the electromagnetic flux that we receive 
from nearby radio--quiet AGNs 
in hard X--rays (see e.g. Tueller et al. 2008).
We now compare this flux with the expected flux of other candidate sources.  
We will consider flaring or bursting sources, that is impulsive events, but the
spreading of the arrival times of UHECRs from a source located at a distance 
$D$, $\Delta t \sim D \theta^2/2 c$, due to even tiny magnetic deflections, 
ensure that we can treat all candidate sources as continuous.
We will estimate the predicted flux in two different ways.

First, assume that a class of sources is characterised by a
pulse of emission of UHECRs, of average total energy $<E>$.  
Assume also that these events occur at a rate $R$ per galaxy, per
year, and consider those events occurring within the GZK radius $D_c$.  We
have:
\begin{equation}
F \, = \, <E> {R \over 3.15 \times 10^7} 
\, { N_{\rm g}(D<D_c) \over 4\pi (a D_c)^2} 
\end{equation}
where $3.15\times 10^7$ is the number of seconds in one year 
and $N_{\rm g}(D<D_c)$ is the number of galaxies within $D_c$
of $L_*$ luminosity.
The average distance of the sources is $a D_c$ ($a=3/4$ for sources
homogeneously distributed).
Setting the mean local galaxy density 
$n_{\rm g}= N_{\rm g}/(4\pi D_c^3/3)= 10^{-2} n_{\rm g, -2}$ Mpc$^{-3}$
we have:
\begin{equation}
F  \sim 1.2 \times 10^{-57}  <E> R\, n_{\rm g, -2} {D_{c, 100} \over a^2}
 \,\,\, {\rm erg\, cm^{-2}\, s^{-1}}
\label{rate}
\end{equation}  
where $D_c =100 D_{c,100}$ Mpc.

The second estimate on the predicted UHECR flux uses the
electromagnetic flux as a proxy.  Assume that we detect, for a typical member
of a class of sources, an average fluence $<{\cal F}>$, and that there are $N$
events per year.  If a fraction $\eta$ of these events comes from sources
within $D_c$, we have
\begin{equation}
F\, = \, {\eta <{\cal F}> N  \over 3.15 \times 10^7} 
\label{m2}
\end{equation}
This estimate is more appropriate when dealing with sources, such as
long and short GRBs, whose fluences and occurrences are known, while Eq.
\ref{rate} is more appropriate when dealing with possible sources of unknown
electromagnetic output, but predicted energetics and rates, such as newborn
magnetars (Arons 2003) or giant flares from old magnetars.

Let us consider the above classes of sources in turn, starting from short GRBs.  
In the BATSE catalog (cossc.gsfc.nasa.gov/docs/cgro/batse/BATSE\_Ctlg/flux.html) 
we have 490 short GRBs of total fluence $5.5\times 10^{-4}$ erg cm$^{-2}$ in 9
years of operation. 
Tanvir et al. (2005) correlated these short GRBs with local optically selected 
galaxies finding that a fraction between 5 and 25\% of BATSE short GRBs might
be nearby, i.e. at $z < 0.025$, corresponding to 109 Mpc.
Considering that BATSE saw half of the sky and setting $\eta=0.1$ we
have an average flux of $3.9\times 10^{-13}$ erg cm$^{-2}$ s$^{-1}$.
%
%
%
%
Then, if the UHECR flux is similar to the electromagnetic one, 
short GRBs do not match the required flux.

Classical long GRBs (namely, energetic GRBs at $z\gsim 1$)
in the BATSE sample have a 
total fluence of 0.024 erg cm$^{-2}$
(for the listed 1490 long GRBs in the BATSE catalog),
corresponding to an average (all sky) flux of 
$1.7\times 10^{-10}$ erg cm$^{-2}$ s$^{-1}$, larger than the one
given by Eq. \ref{fa}.
However, for long BATSE GRBs, $\eta$ must be much smaller than 0.1,
as directly suggested by the paucity of nearby events, and by the lack
of correlation with nearby galaxies and clusters (Ghirlanda et al. 2006).
While we cannot dismiss them as sources of UHECRs, it seems likely
that classical BATSE bursts are too distant (but see below).

%


Consider now giant flares from relatively ``old'' magnetars.  The giant flare
from SGR 1806--20 of Dec 27, 2004 emitted an energy $E\sim 10^{46}$ erg in less
than a second.  The radio afterglow convincingly demonstrated the formation of
a (at least mildly) relativistic fireball.  With the current hard X--ray
instruments, such flares can be detected up to $\sim$30--40 Mpc (Hurley et al.
2005).  Eq.  \ref{rate}, with $D_{c,100}=0.3$, and $<E>=10^{46}$erg, would
require $R\sim 1$ event per galaxy per year, while an approximate limit to the
rate is $R< 1/30$ yr$^{-1}$ (see e.g. Lazzati et al. 2005).

Finally, consider fastly spinning newly born magnetars, whose rotational
energy can exceed $10^{52}$ erg, with a rate of $R=10^{-4}$ events per galaxy
per year (Arons 2003).  
With the estimated galaxy density 
($n_{\rm g, -2}\sim 0.7$ with $L\sim L_*$; Blanton et al. 2001) 
there should be 1 event per year within 100 Mpc.  
If each magnetar produces $10^{50}$ erg in UHECRs, then this class of sources
can be the progenitor of the Auger events (Eq. \ref{rate}).  
This is independent of collimation, since the reduced rate of events pointing 
at us is compensated by an increase of the apparent energetics.
But if an equal amount of energy is released in electromagnetic form, at
energies detectable by BATSE, 
then they should be a significant fraction of all BATSE GRBs. 
Since the birth of a magnetar should be accompanied by a supernova,
{\it these events should be associated with long, rather than short GRBs,}
for which no associated supernova has been seen.
If the radiative output is isotropic, they will all be nearby, sub--energetic, GRBs.

The required fluence of these sub--energetic nearby long GRBs, to match the
UHECRs flux, should be
\begin{equation}
<{\cal F}>\, \sim\,  3.15\times 10^{-4} \epsilon_{\rm CR} \, {F_{\rm A, -11} \over \eta N}
\,\,\, {\rm erg\,\, cm^{-2}} 
\end{equation}
where $\epsilon_{\rm CR}$ is the ratio of the emitted energy in radiation and UHECR.
If $\eta\sim \epsilon_{\rm CR}\sim 1$, these events constitute a 
sizeable fraction of the total fluence of all long BATSE 
GRBs in one year 
(which is ${\cal F} \sim 0.024/9\sim  2.7\times 10^{-3}$ erg cm$^{-2}$).

%

Since we know that the large majority of long GRBs are not nearby, newly
born magnetars should not constitute conspicuous events in hard X--rays.
Their fluence must be mostly emitted in another energy range.
GRB 060218 (Campana et al. 2006) with an energy of a few $\times 10^{49}$ erg,
at a distance of 145 Mpc, could be one of these events, and
Soderberg et al. (2006) and Toma et al. (2007) already suggested
that this GRB was powered by a newly born magnetar.
The spectrum of its prompt emission peaked at $\sim$5 keV, i.e. its fluence
in relatively soft X--rays exceeded the 15--150 keV fluence.
It was also very long, slowly rising, and would not have been detected by BATSE.
Soderberg et al. (2006)
pointed out that these sub--energetic long GRBs should not be strongly beamed
(not to exceed the rate of SN Ib,c), and should occur at a rate of
$230^{+490}_{190}$ Gpc$^{-3}$ yr$^{-1}$, corresponding to 
$R \approx 10^{-5}$ events per $L_*$ galaxy per year, about ten times larger 
than for classical long GRBs whose radiation is collimated into 1\% of the sky.
According to this rate, Eq. \ref{rate} would then demand $<E>\sim 6\times 10^{50}$ erg
in UHECRs to match the observed flux.

\section{Conclusion}

We have correlated the cosmic rays with $E>57$ EeV 
detected by the Auger Observatory with a complete, absorption--free sample of
\hi\ selected galaxies. 
We found a significant
correlation when correlating the \hi\ {\it flux} of galaxies of our sample.

When considering the largest 95HIPASS catalogue and the 27 UHECRs we find a
weak correlation (probability of 72\%), while a larger significance (87.8\%) is
reached if we consider the most complete 99HICAT sample of galaxies (though
with 25 UHECRs). 
These probabilities are maximised by cutting the \hi\ sample in
distance or luminosity bins:
it becomes 99\% when considering the 500 most luminous (or most \hi\ massive)
galaxies (1/4 of the sample), and 98\% when considering the 500 galaxies lying
between 38 and 54 Mpc, where the Centaurus cluster of galaxies is.
 
Thus there is the possibility that the UHECRs coming from the direction
of Cen A are instead coming from the more distant Centaurus cluster.
Galaxies of this cluster are richer in \hi\ than Virgo galaxies,
explaining why there is no UHECR event from the direction of Virgo.

This sample is formed by \hi\ emitting galaxies, therefore it is biased against
ellipticals.  The found correlation with these galaxies, per se, is not
disproving the found correlation with AGNs (Abraham et al. 2007, 2008; George
et al. 2008), since they also trace the local distribution of matter, as
spiral galaxies do.  
On the other hand, it opens up the possibility, on equal
foot, that UHECRs are produced by GRBs or newly born magnetars
(see also Singh et al. 2004 who used AGASA events).
With the
caveat that it is premature, with so few events and big theoretical
uncertainties, to draw strong conclusions, we have pointed out that although
classical (i.e. energetic) long GRBs and short GRBs have difficulties in
producing the required UHECR flux, newly born magnetars can.  If so,
they could also be a subclass of {\it long} GRBs, possibly sub--energetic and
relatively nearby, powered by fastly spinning, newborn magnetars.  
The future increased statistics of UHECRs arrival directions will help
to discriminate among the different proposed progenitors, especially if there
will be (or not) an excess of events close to the radio core and/or lobes of Cen A.


\section*{Acknowledgments}
We thank the referee for constructive comments.
We thank the ASI I/088/06/0 and the 2007 PRIN--INAF grants and
Ivy Wong and Martin Zwaan for providing the NHICAT catalogue.
The Parkes telescope is part of the Australia Telescope which is funded by 
the Commonwealth of Australia for operation as a National Facility managed by CSIRO.


\begin{thebibliography}{}
\bibitem[]{} Abbasi R.U., et al., 2008 preprint (astro--ph/0804.0382)
\bibitem[]{} Abraham J., et al., 2004, Nucl. Instrum. Methods A, 523, 50
\bibitem[]{} Abraham J., et al., 2007, Science, 318, 938
\bibitem[]{} Abraham J., et al., 2008, Astroparticle Physics, 29, 188
\bibitem[]{} Abbu--Zayyad T., et al., 2000,  Nucl. Instrum. Methods A, 450, 253
\bibitem[]{} Arons J., 2003, ApJ, 589, 871
\bibitem[]{} Barnes D. G., et al., 2001, MNRAS, 322, 486
\bibitem[]{} Blanton M.R., et al. 2001, AJ, 121, 2358
\bibitem[]{} Borkowski J., Gotz D., Mereghetti S., Mowlavi N., Shaw S., Turler
             M., 2004, GCN Report No. 2920
\bibitem[]{} Campana S., et al., 2006, Nat, 442, 1008
\bibitem[]{} Cannizzo J.K. et al., 2006, GCN rep. 20.1 
\bibitem[]{} Dermer C.D., 2007, preprint (astro-ph/0711.2804)
\bibitem[]{} Fasano G. \& Franceschini A., 1987, MNRAS, 225, 155
\bibitem[]{} George M.R., Fabian A.C., Baumgartner W.H., Mushotzky R.F. 
             \& Tueller J., 2008, MNRAS, 388, L59
\bibitem[]{} Ghirlanda G., et al., Magliocchetti M., Ghisellini G., Guzzo L., 
              2006, MNRAS, 386, L20
\bibitem[]{} Gorbunov D., Tinyakov P., Tkachev I. \& Troitsky S., 2008,
             preprint (astro--ph/0711.4060)
\bibitem[]{} Harari D., Mollerach S. \& Roulet E., 2006, JCAP, 11, 12 
\bibitem[]{} Hartman R.C., et al., 1999, ApJS, 123, 79 
\bibitem[]{} Hurley K. et al., 2005, Nat, 434, 1098
\bibitem[]{} Lazzati D., Ghirlanda G. \& Ghisellini G., 2005, MNRAS, 362, L8
\bibitem[]{} Meyer et al. 2004, MNRAS, 350, 1195
\bibitem[]{} Milgrom M. \&  Usov V., 1996, ApJ, 449, L37
\bibitem[]{} Moskalenko I.V., Stawarz L., Porter T.A. \& Cheung C.C., 2008,
             subm to ApJ (astro--ph/0805.1260)
\bibitem[]{} Murase K., Ioka K., Hagataki S \& Nakamura T., 2008, PRD, 78, id. 023005
             (astro--ph/0801.2861)  
\bibitem[]{} Nagar N.M. \& Matulich J., 2008, A\&A in press (astro--ph/0806.3220)
\bibitem[]{} Ohoka H., et al., 1997, Nucl. Instrum. Methods A, 385, 268 
\bibitem[]{} Peacock J.A., 1983, MNRAS, 202, 615
\bibitem[]{} Singh S., Ma C.-P. \& Arons J., 2004, Phys. Rev. D, 69, 063003
\bibitem[]{} Soderberg A.,  et al., 2006, Nat, 442, 1014  
\bibitem[]{} Sommers P., 2001, Astroparticle Physics, 14, 271
\bibitem[]{} Staveley--Smith L., et al., 1996, Publ. Aston. Soc. Aust., 13, 243
\bibitem[]{} Tanvir N. R., Chapman R., Levan A. J., Priddey R. S., 2005, Nat, 438, 991l
\bibitem[]{} Terasawa T., et al., 2005, Nat, 434, 1110
\bibitem[]{} Toma K., Ioka K., Sakamoto T. \& Nakamura T., 2007, ApJ, 659, 1420 
\bibitem[]{} Torres D.F. \& Anchordoqui L.A., 2004, RPPh, 67, 1663 (astro--ph/0402371)
\bibitem[]{} Tueller J., Mushotzky R.F., Barthelmy S., Cannizzo J.K., Geherels N., Markwardt C.B. 
             \& Winter L.M., 2008, ApJ, 681, 113
\bibitem[]{} V\'eron--Cetty M.-P., \& V\'eron P., 2006, A\&A 455, 773
\bibitem[]{} Vietri M., 1995, ApJ, 453, 883
\bibitem[]{} Wang X--Y., Razzaque S. \& M\'esz\'aros P. 2008, ApJ, 677, 432
\bibitem[]{} Wong O.I., et al.  2006, MNRAS, 371, 1855
\bibitem[]{} Waxman E., 1995, Phys. Rev. Lett., 75, 386
\bibitem[]{} Zwaan M.A., et al., 2004, MNRAS, 350, 1210

\end{thebibliography}
\end{document}